\documentclass[twocolumn]{aa}
\usepackage{graphicx}
\usepackage{txfonts}
\begin{document}
   \title{Restoring Color-magnitude Diagrams with the 
Richardson-Lucy Algorithm}

   \subtitle{}

   \author{M. Cignoni
          \inst{1,2}
          \and
          Steven N. Shore \inst{1,3}}
 \offprints{M. Cignoni}

   \institute{
(1) Dipartimento di Fisica ``Enrico Fermi'', 
Universit\`a di Pisa, largo Pontecorvo 3, Pisa I-56127 Italy;\\
(2) INAF - Osservatorio Astronomico di Capodimonte, Via Moiariello 16,
     80131 Napoli, Italy;\\
(3) INFN - Sezione di Pisa, largo Pontecorvo 3, Pisa I-56127,
     Italy}

%DRAFT 5/2/2006

   \date{Received; accepted }

% \abstract{}{}{}{}{} 
% 5 {} token are mandatory
 
  \abstract{We present an application of the  Richardson-Lucy
algorithm to the analysis of color-magnitude diagrams by converting
the CMD into an image and using a restoring point spread function
function ({\it psf}) derived from the known, often complex, sources of
error. We show numerical experiments that demonstrate good recovery
of the original image and establish convergence rates for ideal cases
with single gaussian uncertainties and poisson noise using a $\chi^2$
statistic.  About 30-50 iterations suffice.  
As an application, we show the results
for a particular case, the  Hipparcos sample of the solar
neighborhood where the uncertainties are mainly due to parallax which
we model with a composite weighted gaussian using the observed error
distributions.  The resulting {\it psf} has a slightly narrower core
and broader wings than a single gaussian.  The reddening and
photometric errors are considerably reduced by restricting the sample
to within 80 pc and to $M_V \leq 3.5$. We find that the  recovered
image, which has a narrower, better defined main sequence and a more
clearly defined red giant clump, can be used as input to stellar
evolution modeling of the star formation rate in the solar vicinity
and, with more contributing uncertainties taken into account, for
general Galactic and extragalactic structure and population studies.

   }

   \maketitle

%
%________________________________________________________________

\section{Introduction}
The color-magnitude  diagram (CMD) 
is the key tool for studying evolution in stellar populations. 
However, when interpreting cluster and field
CMDs we have to face different problems. For clusters, unless they are very
nearby and large, the stars can be assumed to be at nearly the same distance
and therefore even if the absolute luminosities are unknown their relative
values are well determined.  Unless the system is very young, of order
the pre-main sequence contraction time, the stars can be assumed to
have the same age and metallicity since their formation occurs essentially
instantaneously. The uncertainties are, consequently, mainly photometric
and also due to interstellar reddening (but this, like the distance, can
initially be assumed identical for all members) or in crowded fields
due to confusion.  In contrast, for field stars, the uncertainties
affecting the CMD are mainly due to parallax errors and 
reddening, which are different for each star, so the
relative as well as absolute luminosities are inherently uncertain.
Since the aim of field studies is to determine the structure and
the star formation and metallicity histories of the solar
neighborhood, any sample must be assumed to have 
formed over a long time.  To unravel these histories requires a
different approach than usually adopted for the treatment of the
aggregate CMD.  Our aim in this study is to show how to obtain a
cleaned input for such analyses that is, in effect, a restoration of
the intrinsic distribution to the limit of the errors.  
A CMD may be regarded as an image, the 'intensity' 
being the number of stars 
in a bin of some photometric indices or effective temperature and luminosity, 
affected by a point
spread function ({\it psf}) that originates from the error distributions of 
the parallaxes, interstellar reddening, and photometry and 
we propose using the same techniques that have been developed for image
restoration.  

Specifically, we apply the Bayesian Richardson-Lucy algorithm
(Richardson \cite{richardson}, Lucy \cite{lucy74}) to an observed CMD.  
This method is very well known in the astronomical
community and has been broadly applied to recover data from images and
histograms (see, e.g. Bertero \& Boccacci \cite{BB} and references
therein). Here we suggest another use, this time for the study of
stellar populations.  If we grid a CMD by building a two dimensional histogram
in, say, $M_V$  and $B-V$, the data is an image blurred by a {\it psf}
that is the matrix produced by the 
observational errors.  The analysis then becomes a
deconvolution problem.\footnote{ There is, however, a difference with
  respect to a real astronomical image.  There is no ``background'',
  or if it is due to contamination of the sample by stars at very
  different distances this can be modeled within the kernel.  The
  errors here are strictly poissonian as a result of the binning.}   
From this point of view, the loss of
information about single stars is compensated by the opportunity to
analyze the sample data in a statistically consistent 
sense using imaging methods.  
Our intent is to recover, as closely as possible within the limits of
the {\it psf}, the intrinsic CMD for comparison with stellar
isochrones that may then be analyzed by any of many available statistical
methods (see especially Tolstoy \& Saha \cite{tol}, Hernandez et al. 
\cite{her99},\cite{her00}). Although in what follows we use the Hipparcos
solar neighborhood sample for illustration,\footnote{The technique was
  developed as part of a study of the star formation history of the
  solar neighborhood (Cignoni \cite{cignphd}) and a more complete study of the
  results of its application is in preparation.} the technique is as
general as the RL algorithm itself and we emphasize that our aim is to
propose a new view of how to treat observatioanl CMD 
data for stellar and galactic
evolution.

\section{Use of the algorithm}

Bayes' theorem provides the relationship between the probability of a
hypothesis (model parameters) conditional on a given data, $P(M|D)$ 
(posterior), and the probability
of the data conditional on the hypothesis, $P(D|M)$, and the prior
probability $P(M)$:
\begin{eqnarray} 
&&P(M|D)=\frac{P(D|M) P(M)}{\int P(D|M)P(M)}\label{mid1}
\end{eqnarray}
where $M$ are the values for the parameters of the model (the hypothesis)
and $D$ are the measured data.  The 
prior is the probability of any hypothesis being true 
before empirical data, the posterior probability is what we can know about
hypothesis after the observation (how likely, given the observations,
our model is as an explanation of the data); $P(D|M)$, the likelihood,
is the chance of the observation $D_n$ given the model $M_n$ for
some sample $n$ (e.g. Jaynes 2003).  First, assuming the intrinsic
noise in the data is poisson distributed, the conditional 
probability is given by:
\begin{eqnarray} 
&&P(D|M)=\prod^{N}_{n=0}\frac{\exp{(-{M}_n){{M}_n}^{D_n}}}{{D_n}!}
\end{eqnarray}
where the product is taken over N bins.  The intrinsic fluctuations
are poisson distributed.  

The {\it psf} is from the photometric and parallax errors.  
The data value is the number of stars in
the $nth$ bin which result from sampling some initial mass function
(IMF) $\phi(m)$ at a star formation rate (SFR), $\psi(t)$, 
and depending on some
metallicity history, $Z(t)$; if we fix $\phi$ the other two are the
functions should be recoverable from the sample.  
The relevance to CMDs is that, depending on the type of data, there
may be several contributing factors to the total {\it psf}, $K$ 
e.g. errors in photometry, parallaxes, and proper motions, and
uncertainties in reddening, unresolved binary companions, etc..  The 
combination of blurring and fluctuations on the final image is:
\begin{eqnarray}  
({\rm image}) =K \star ({\rm object}) + {\cal N}
\label{frd3}
\end{eqnarray}
where ${\cal N}$ is the noise.  Richardson (\cite{richardson}) and 
Lucy (\cite{lucy74}) were the first to propose an iterative inversion
algorithm (hereinafter R-L algorithm) for deconvolution:
\begin{eqnarray}  
&&f^{(k+1)}=f^{(k)}\,K\star \left(\frac{D}{K\star f^{(k)}}\right)
\end{eqnarray}
with $f^{(k)}$ is the object estimation 
after the k-th iteration and $D$ is the data.
The method is derived assuming poisson noise and taking a constant 
prior information $P(M)$ in equation \ref{mid1}
(non-informative prior). It can be shown that the algorithm converges to a
restoration that when re-blurred by the {\it psf} 
is close to the data in a maximum
likelihood sense (for details see e.g. Shepp \& Vardi \cite{shepp}).  

Although the Hipparcos data is the subject 
of our application of the R-L algorithm, 
its possible use extends to {\it any} binned CMD.  Since the 
specific deconvolution problem must be preceded by a careful analysis of the
associated uncertainties, the following section discusses the sources of
blur in the Hipparcos data.  In the following sections we show tests 
based on convolving synthetic CMDs, mimicking the solar
neighborhood Hipparcos sample, with different {\it psf}s and performed 
restorations using the R-L method.  We show examples of experiments
that demonstrate successful image restoration under various assumed
point spread functions.  The same analysis was performed after adding poisson
noise and using the $\chi^2$ statistic as the convergence parameter.  The last 
section shows the result of applying the method to the {\it real} data
using an observationally derived {\it psf} and discuss some
applications.

\section{Hipparcos uncertainties}
Our sample was selected using two criteria. The stars are all 
brighter than $V \sim 8$, which is comfortably within the completeness 
interval assumed by the Hipparcos collaboration of between 7.3 and 9
mag depending on Galactic latitude and color (e.g. Perryman et al. 
1995), and are within 80 pc from the Sun, 
  corresponding to an absolute magnitude of $3.5$ mag at our adopted 
limit.  The parallax precision is generally better than 10\% and the
nonlinearity bias is assumed very small (e.g. Arenou \& Luri \cite{arenou2}).  
The histogram of the absolute visual magnitude error distribution for our
sample, obtained by propagating 
the parallax errors, (see fig. 1), shows a mean error 
of about 0.10 mag with a standard deviation of about 0.05 mag. 
The same figure shows the absolute magnitude error 
distribution for stars in three
bins of absolute visual magnitude, respectively $2-2.5, 2.5-3, 3-3.5\,\,
\mathrm{mag}$. The histograms for stars with $2 < M_{V} < 2.5$ and $2.5 <
M_{V} <3$ are the same within the poisson 
fluctuations, 
while the histogram $3-3.5$ presents a slightly systematic shift
(about $0.01-0.02$ mag) toward bigger errors.  This last feature may
be due to the selection in absolute magnitude. 
Moreover, the parallax precision 
decreases with distance so the larger absolute magnitude errors found in the
$3-3.5$ mag could be explained by a greater mean distance of the
sample. It is, however, a small effect so the distribution for the absolute
magnitude error, to fair accuracy, was assumed to be 
independent of the position in the CMD 
as a first assumption about the {\it psf}.
\begin{figure}
\centering
\centering \includegraphics[width=7cm]{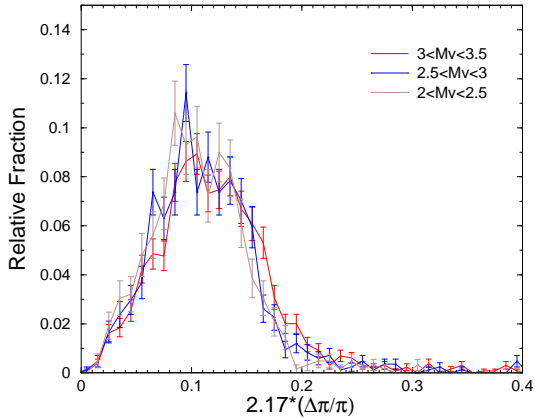}  
\caption{The absolute 
magnitude error distribution for Hipparcos 
stars brighter than $M_V = 3.5$ and nearer than 80 
pc for different intervals of absolute 
magnitude. The error bar is poissonian.  
The figure shows the general agreement among the curves.} 
 \label{nmva}
\end{figure}
The photometric error for stars
brighter than $M_V = 3.5$ is generally $< 0.02$ mag but there 
is also an high error tail due to giant and clump stars.  These last
groups comprise only $1-2\%$ of the total sample and since the 
color uncertainty is so small, compared to the 
absolute magnitude error, we here assume that it can be neglected.
The method we'll present can, however, be easily generalized to two
dimensions (for instance in treating crowding and other photometric
errors in observations of fields in resolved galaxies).

\section{Tests with artificial data}

\begin{figure*}
   \centering
   \includegraphics[width=6cm]{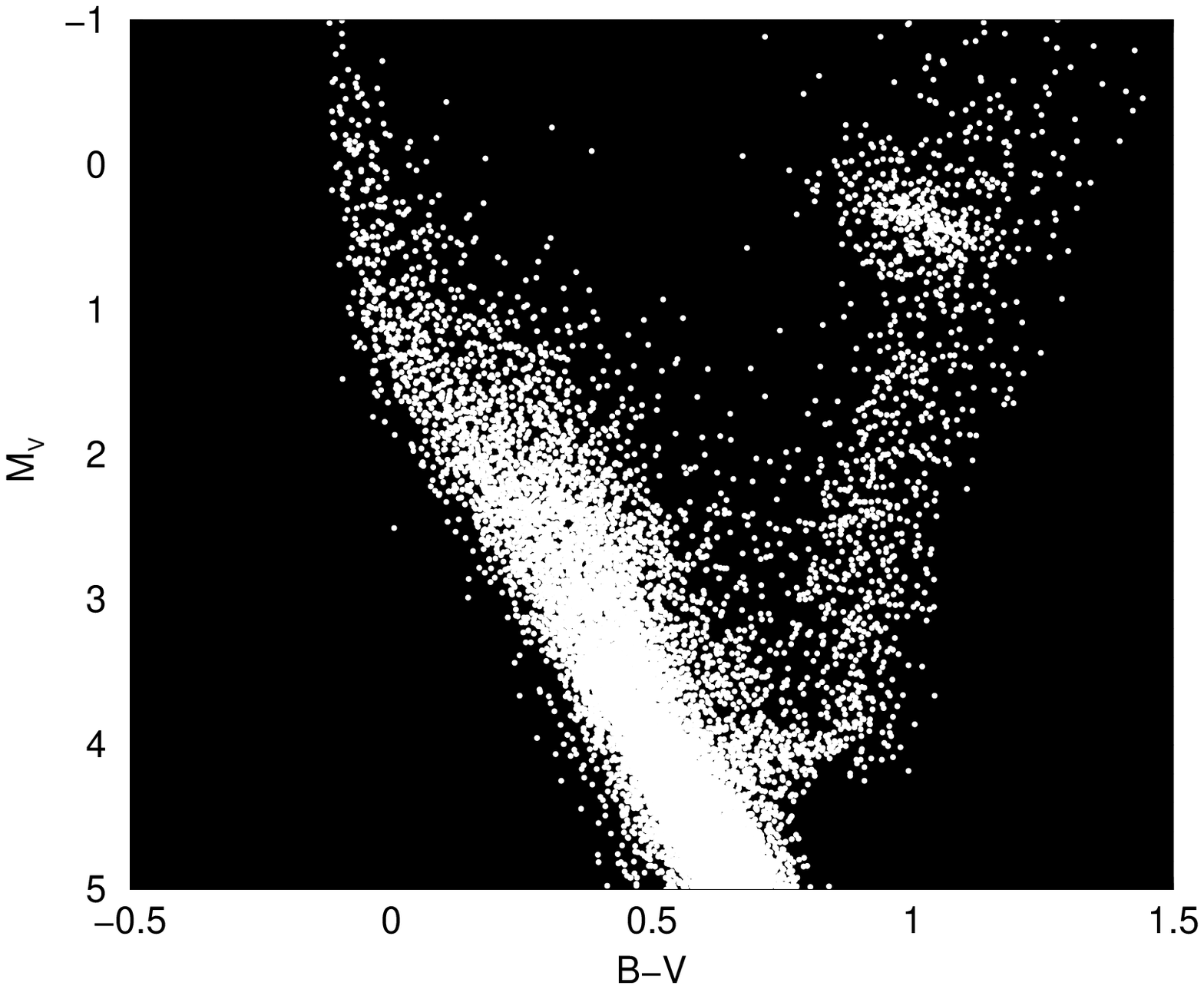}
   \includegraphics[width=6cm]{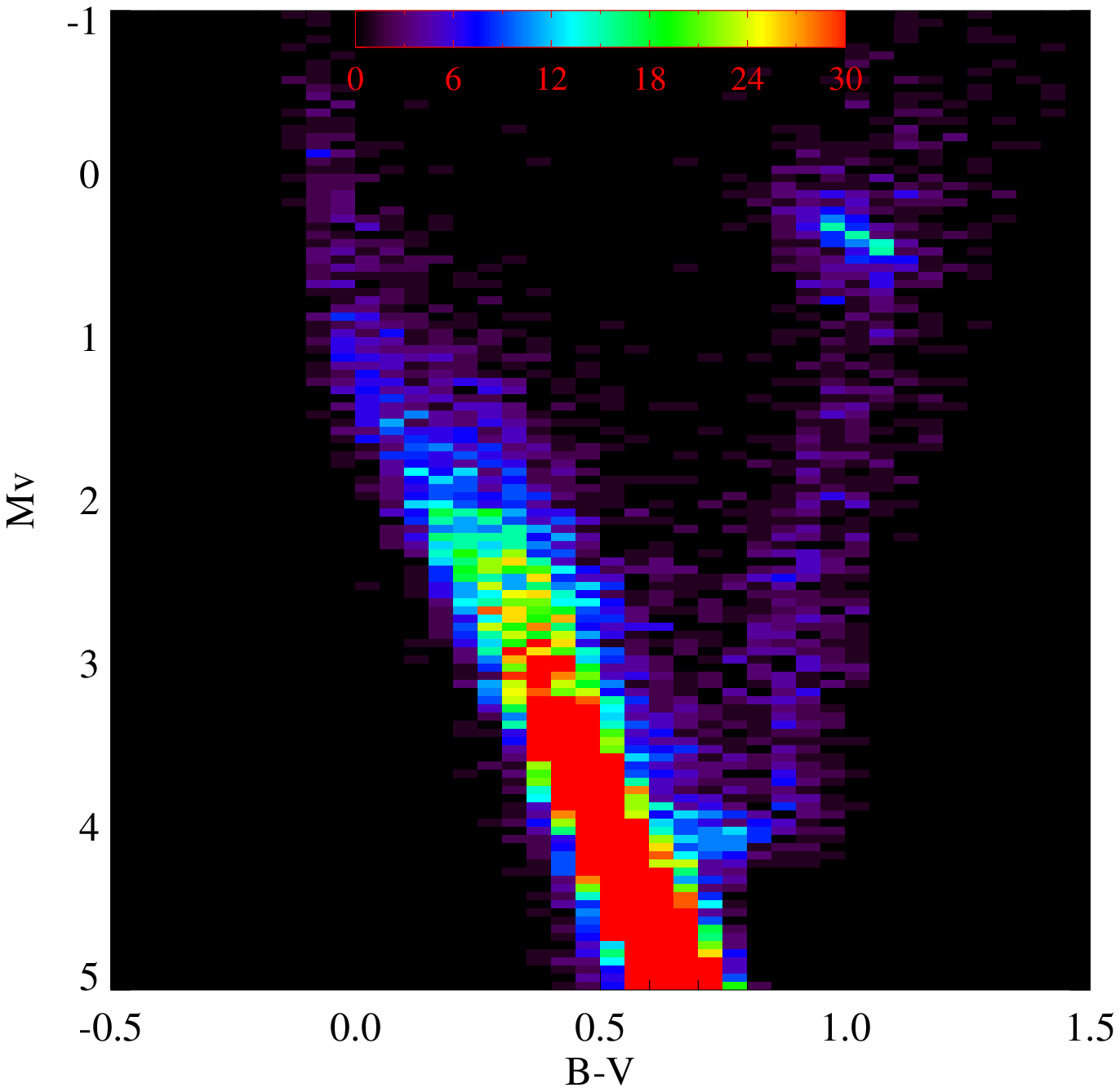}\\
   \includegraphics[width=6cm]{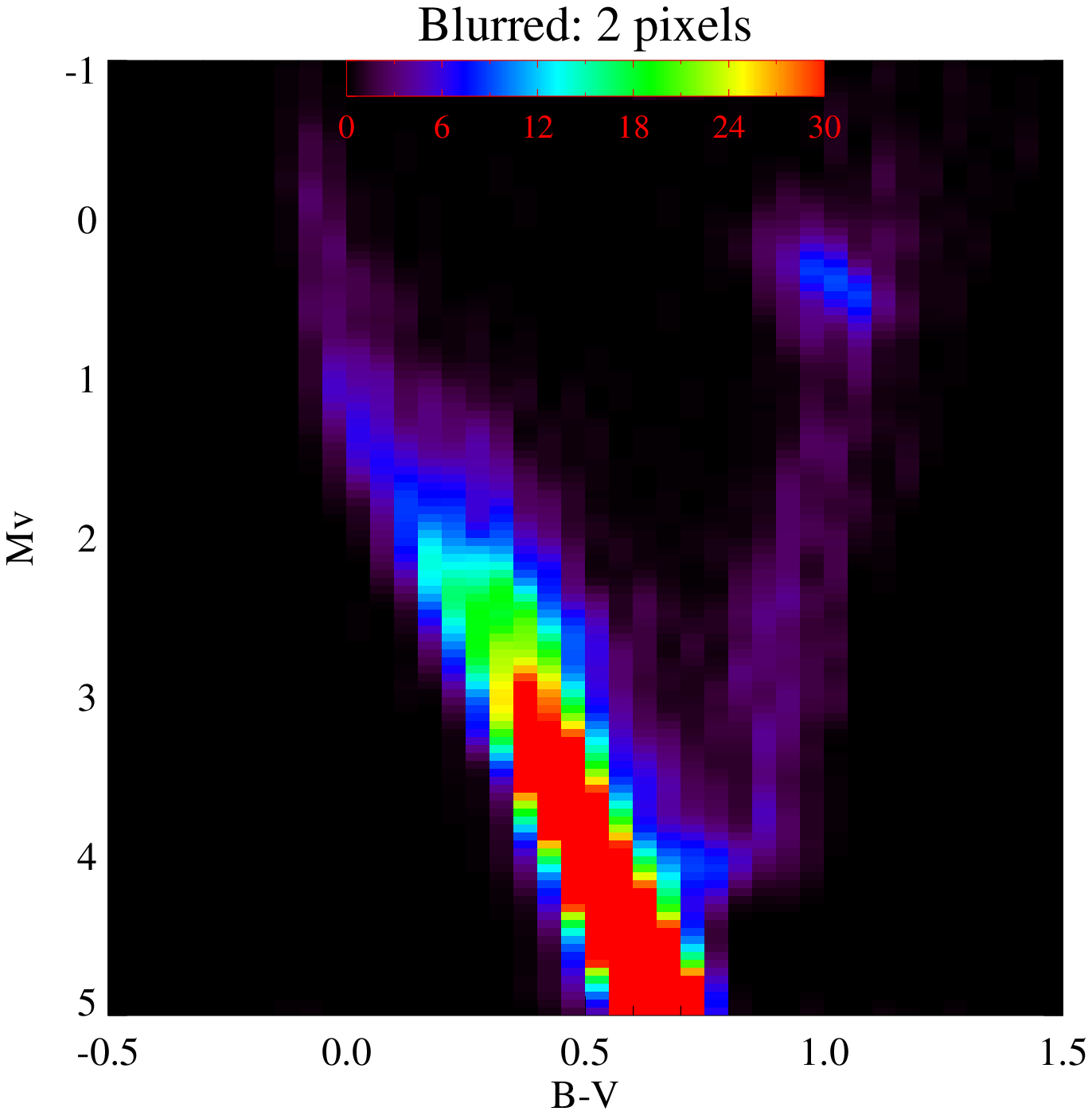}
   \includegraphics[width=6cm]{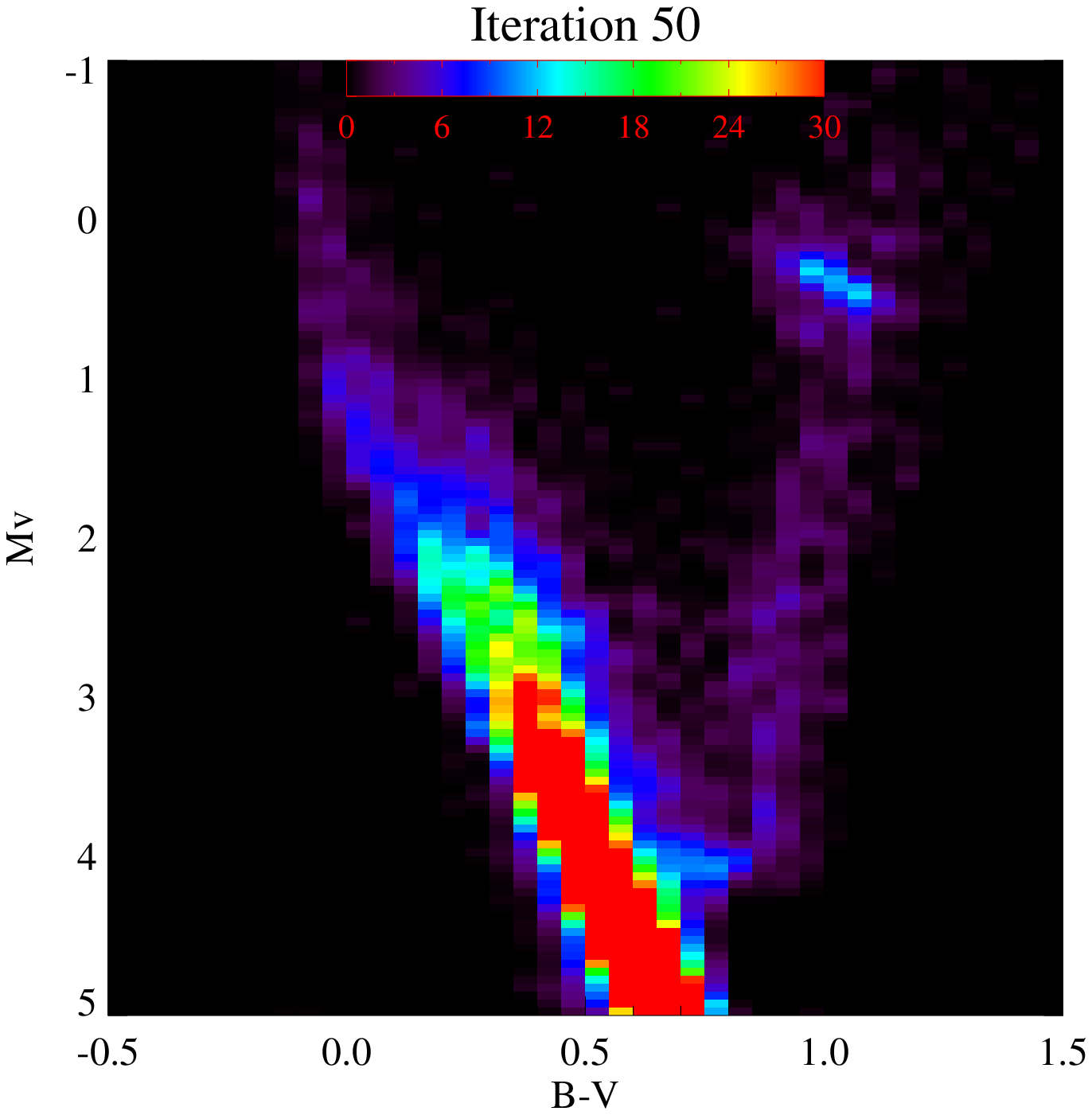}\\
   \includegraphics[width=6cm]{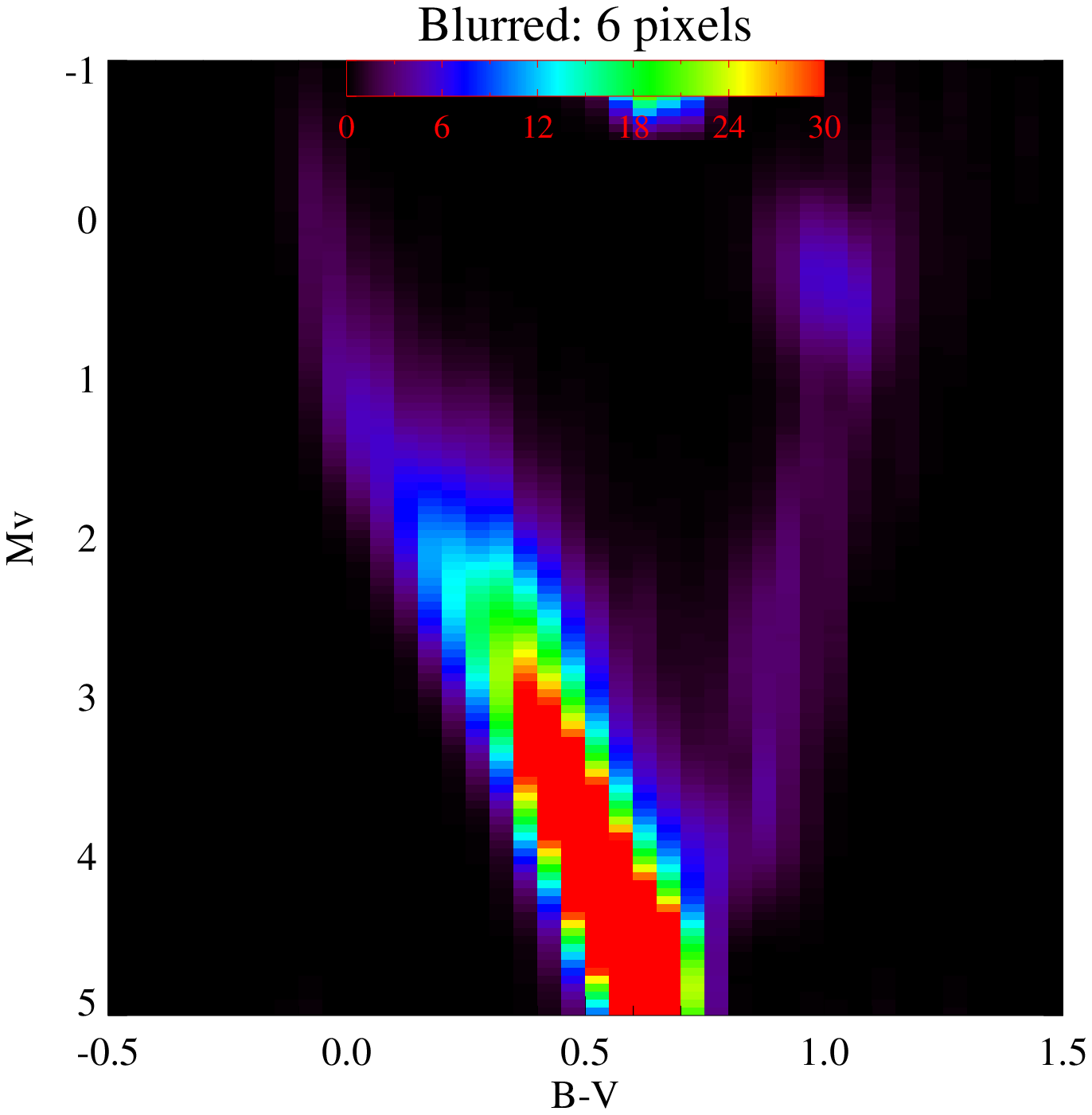}
   \includegraphics[width=6cm]{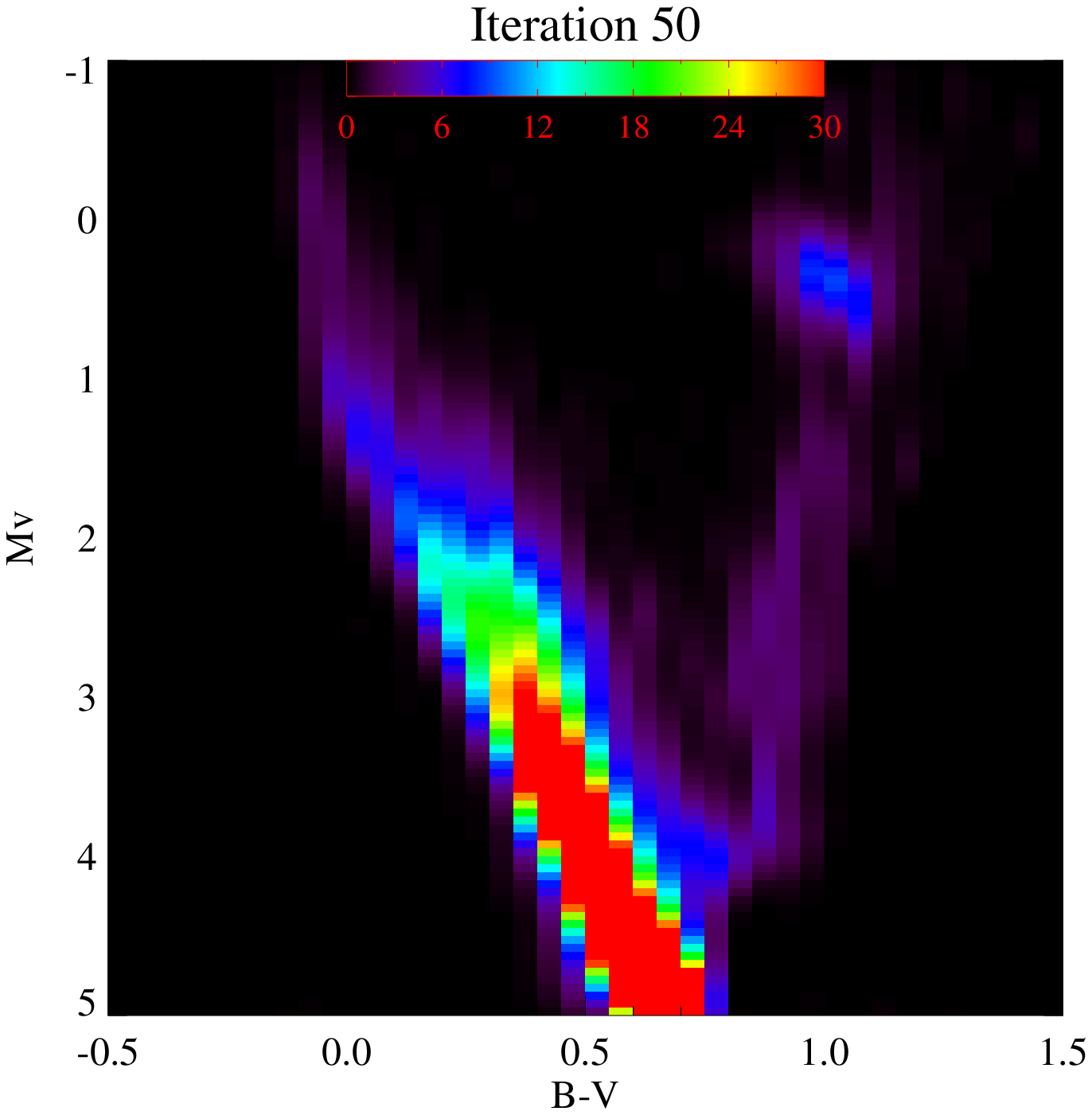}\\
   \caption{Synthetic tests demonstrating the recovery achieved using the R-L
   algorithm.  From top left: (a) Synthetic CMD, (b) 2-D histogram for the
   synthetic CMD; (c) blurred with 2 pixels gaussian {\it psf}; (d) restored image
   after 50 iterations.  For contrast, (e) synthetic image blurred with 6
   pixels gaussian; (f) restored image for 6 pixels blurring after 50
   iterations (see text for details). }
   \label{matrix}%
    \end{figure*}

Before using our method to real data, we have to test its ability in restoring
CMDs on the basis of artificial data. 
We begin by modeling the {\it psf} with a  gaussian. Considering the  relative
uncertainties in color and absolute magnitude 
we chose $\sigma = 0.1\, \mathrm{mag}$, 
the mean value found in the observational distribution. 
The error on $B-V$ is negligible so the convolution and 
deconvolution affect only the absolute magnitudes but the method can
be generalized to two dimensions.  
The synthetic CMD (fig. 2-a) was 
generated with a Monte Carlo technique (see for details Castellani et
al. \cite{cast} and Cignoni et al. \cite{cign}) using a power law with
a Salpeter exponent (see Kroupa \cite{krou}), a flat 
SFR ($\psi = {\rm constant}$) 
and an observational age-metallicity relation $Z(t)$ from 
N\"ordstrom et al. \cite{nord}. The artificial stars were
placed on grid of tracks (Pisa 
Evolutionary Library\footnote{http://astro.df.unipi.it/SAA/PEL/Z0.html}).  No
binaries have been included but this is not an inherent limitation if
appropriate estimates are available for their contributions
(e.g. Hurley \& Tout \cite{hur}).  
The synthetic CMD contains approximately the same number of
stars brighter than $M_V=3.5$ of the selected Hipparcos CMD, 
so we can avoid possible differences due
to the statistical fluctuations. Figure 2-a shows the
synthetic CMD and Fig. \ref{matrix}-b shows the pixelated version, 
binned with a step size of $0.05$ mag in both coordinates so the
{\it psf} has a $\sigma$ of 2 pixels.  
This choice of the bin size is a practical
compromise: smaller values produce statistical fluctuations too large which
could be a problem because noise amplification is a real drawback with 
iterative algorithms. Larger bins reduce or eliminate the
effectiveness of the restoration, reducing also the potential
information content of the dataset.   In our case, the observational
uncertainty in $B-V$ is negligible compared to the mean error in the absolute
 magnitude so we did not test the effects of blurring in color and 
do not need any finer binning.  Choosing 0.05 mag for
 the bin size minimized the problems mentioned above. 
      \\

Convolving the synthetic CMD with a gaussian ($\sigma=2$ pixels)
produces the image shown in figure 2-c. 
The result after 50 iterations is
shown in figure 2-d.  The restored image is quite 
close to the original. To quantify this match we needed a convergence 
criterion (a rule that stops the algorithm when further iterations
do not significantly improve the result).  
Since the original papers it
has been convenient to 
take  $\chi^2$ as this control parameter 
(see Richardson \cite{richardson}, 
Lucy \cite{lucy74}, Bi \& B\"orner \cite{bi}, Molina et al. \cite{molina}): 
\begin{equation}
\chi_k^2=\frac{(\overline{g}-g^k)^2}{g^k}
\label{stopping}
\end{equation}
where $\overline{g}$ is the blurred image and $g^k$ is the estimate of
the observed image resulting after the $k$th iteration.
Calculating the $\chi_k^2$ for each iteration we found a monotonically
decreasing trend with an asymptotic value close to zero. This is
because the synthetic CMD has been convolved with a {\it psf} without
additional noise so R-L algorithm can perfectly recover the original CMD.  
To gain a better understanding of how the algorithm converges, we tried a
broader {\it psf} with $\sigma=6$ pixels. 
The blurred image and the restoration 
results are shown in figures 2-e-f.  

\begin{figure}
\centering 
\includegraphics[width=7cm]{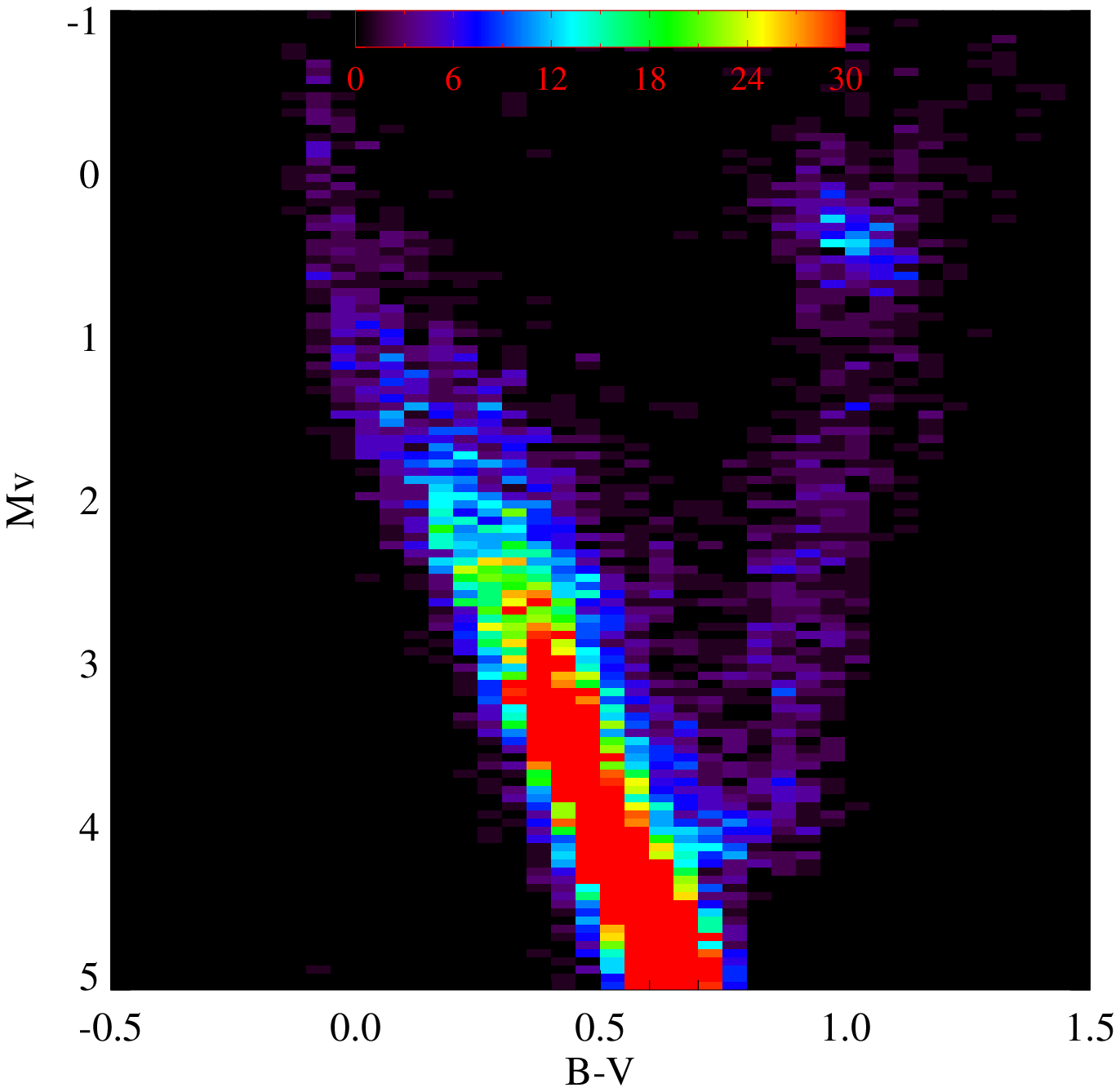}
\includegraphics[width=7cm]{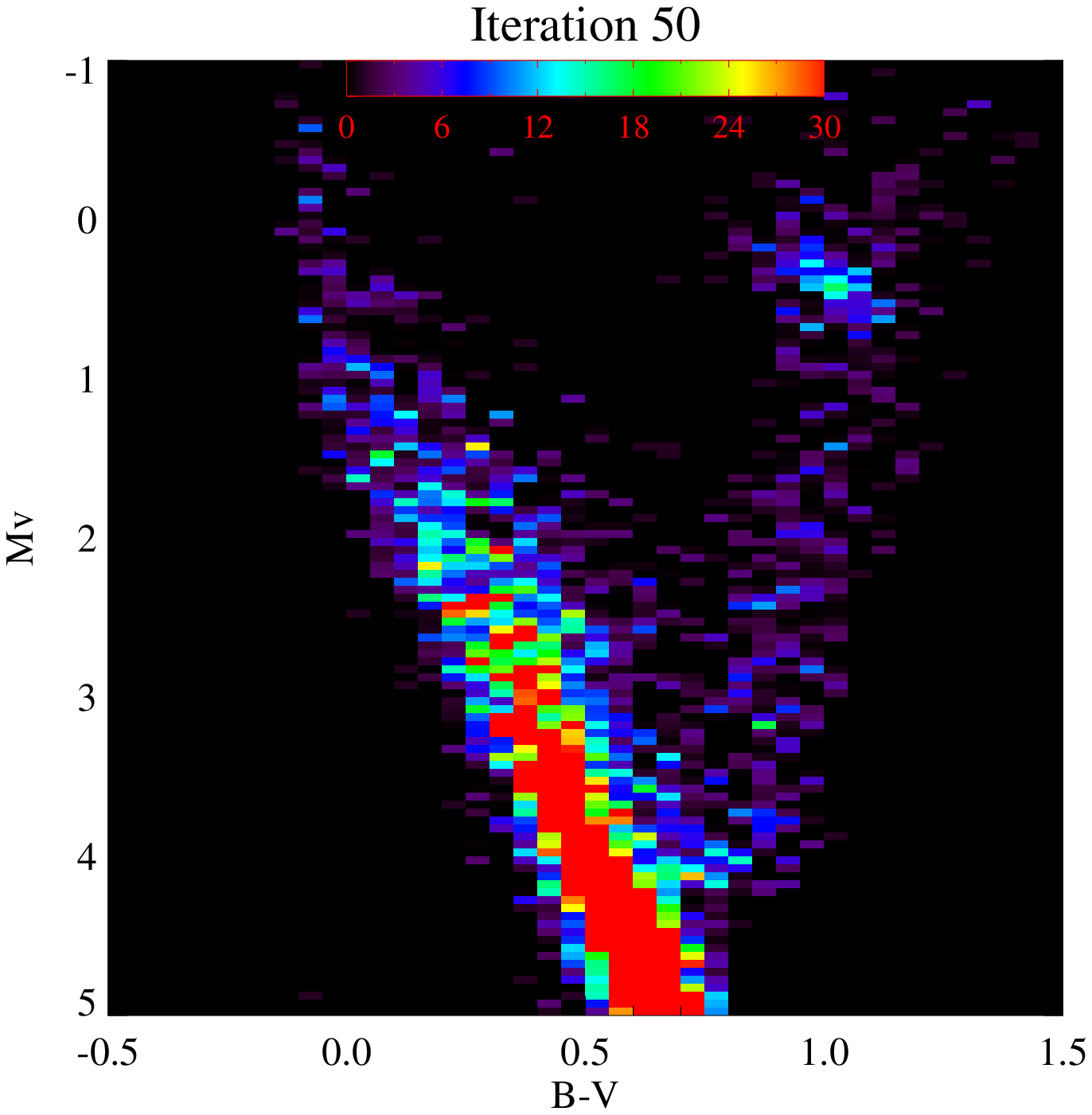}
\caption{(a)  2 pixels blurring plus poisson noise; 
(b) restored CMD after 50 iterations. } 
\label{pois} 
\end{figure}

\begin{figure}
\centering 
\includegraphics[width=7cm]{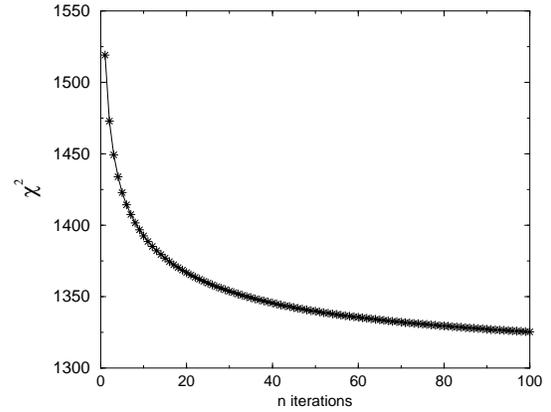} 
\caption{$\chi^2$ versus iterations number (case with blurring and noise).}
\label{poichi}
\end{figure}

Thus far we have treated the case of ideal data (i.e. without noise).  
However, an important point when trying to recover the underlying image from 
noisy data is the sensitivity of the restoration algorithm to noise
and its tendency to produce artifacts with successive iterations.  
The problem is that any convolution is a smoothing operation and, in
the Fourier domain, this reduces the high spatial 
frequencies that characterizes the small scales of the image. Methods like the
R-L  algorithm attempt to remove the smoothing, 
but this is not straightforward since noise is also present.  
As originally suggested by Lucy (\cite{lucy74}), 
the estimate $f^k$ of the real
distribution does not converge as $f^{k+1}-f^k\rightarrow 0$ as
$k\rightarrow \infty$; in fact, after a best estimate of $f^{\tilde{k}}$ the
agreement get worse and artifacts may appear because of small scale
fluctuations.  Since the R-L algorithm is not regularized, when the 
iteration number 
increases noise amplifies in the solution.  In practice, 
the iteration number must be limited to find an acceptable
compromise between resolution and stability.  
To test how the statistical uncertainties present in the data can influence
the deconvolved image, we blurred a CMD with a {\it psf} 
with a $\sigma = 2$ pixels and added poisson noise (see figure
\ref{pois}-a).  The restored CMD after 50 iterations is shown in
figure \ref{pois}-b.
As expected, the result is very different 
from the case from the previous case: the addition
of poisson noise prevents the recovery of the {\it exact} morphology of
the original CMD. In fact, with continued iterations 
the CMD becomes grainier.  The result is better for the most populated regions 
of the CMD, those with  high signal to noise 
ratio, where the CMD features are not buried
by noise.  We find that the 
$\chi^2$, after $\sim 50-100$ iterations, reaches a plateau (see figure
\ref{poichi}) and stabilizes around this value.  This actually
illustrates the utility of the image approach since such changes can
be easily modeled in a controlled way and the changes can be
statistically monitored in a way that bootstrapping, for instance,
cannot show.
In order to preserve the CMD morphology, since the algorithm improves the
statistical properties of the sample but destroys local information,
the iterations are stopped after about 50 cycles 
when the bulk of the restoration has been achieved.  
\section{Application to real data}
\begin{figure}
\centering\includegraphics[width=7cm]{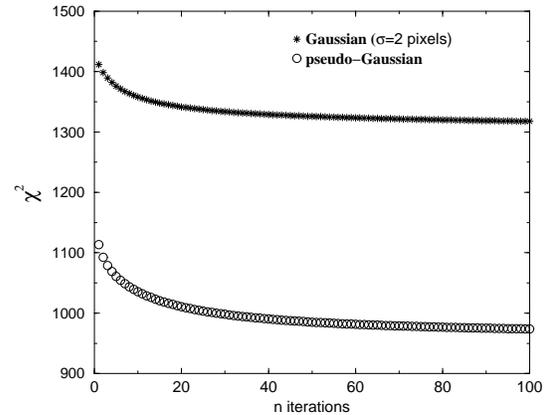}
%\vspace{-0.6cm}
\caption{$\chi^2$ versus iteration number for the restoration with
  gaussian {\it psf} (star) and pseudo-gaussian {\it psf} (filled circles).}
\label{1gau}
\end{figure}

Passing to real data, the first problem is to account for the 
distribution of parallax errors.  In 
principle, we cannot adopt a mean error value as we did in the previous
simulations ($\sigma=2$ pixels).  
If we do not take this point into account and blindly apply the algorithm with
a $\sigma=2$ pixels wide gaussian, the restorations give the $\chi^2$ values
shown in figure \ref{1gau} (star symbol).
A more realistic approach is to use the full information contained in the
data error distribution by forming a {\it psf} from a linear
combination of gaussians using the data shown in Fig. \ref{nmva}:
\begin{equation}
K=\frac{1}{A}\sum_{i}\frac{1}{H_i}\,\frac{1}{\sigma_{i}\sqrt{2\pi}}\exp \left(\frac{-x^2}{2\sigma_{i}^2}\right)
\label{gaussians}
\end{equation}
where the $A$ factor is a normalization constant, 
the weights $H_i$ are the histogram values from figure \ref{nmva}, i.e. 
the fraction of stars with 
absolute magnitude error between $(\Delta M_{V})_i$ and $(\Delta
M_{V})_i+\delta$), and the $\sigma_i$ are the values $(\Delta M_{V})_i$ . 
The resulting {\it psf} (see figure 6) is nearly 
symmetric with a narrower core and broader wings than a gaussian with 
the same full width, and with the dispersion given 
by $\sigma = \sum_j^{} \sigma_j/H_j$, the weighted histogram. 
\begin{figure}
\centering \includegraphics[width=6cm]{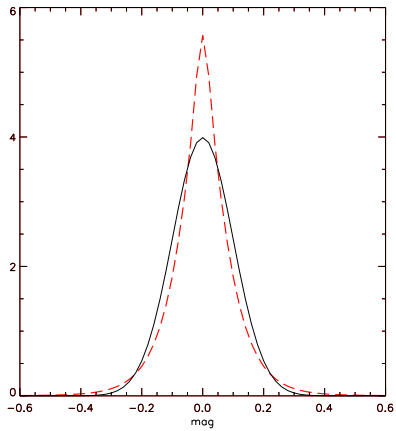}
\caption{The solid line is a gaussian function with $\sigma=2$ pixels.
  The dashed line represents the function obtained from a linear
  combination of gaussians with standard deviations and weights given by data.}
\label{2gau}
\end{figure}
We obtain the sequence of restorations shown in figure \ref{nmv11}
using this composite {\it psf}. Now the $\chi^2$ 
converges faster (see the curve with filled circles in figure
\ref{1gau}) to smaller values so the iterations can be stopped sooner.  
The main results of the restoration process is to compact the CMD features
along the deconvolution axis (absolute magnitude axis).  We find that
(a) the red clump region is compressed and new features appear, (b) 
the giant and sub-giant regions are both better defined, and (c) 
 the main sequence blue edge is now sharper.  These improvements 
in the CMD may appear small, but the restored CMD is the cleanest
restored data we expect to find that is consistent with the
uncertainties in a Bayesian sense.\footnote{As an alternate
  stopping criterion, we suggest dividing the CMD into specific
  critical regions and using a weighted $\chi^2/N$ value with
  the degrees of freedom $N$ depending on region.  For instance, a
  region that is empty and remains always empty need not be included
  in the total number for the reduced $\chi^2$.  A 
  possible weighting for CMD analyses that includes extra
  information coming from the astrophysical setting is to weight the
  regions by their relative ``importance''.  Since the most
  information (in a statistical sense, Jaynes 2003) comes 
from the least probable event, this can be incorporated into the
convergence criterion.  The 
relative probability of finding a star in any part of the CMD for a
  cluster {\it or} the field sample depends on the relative rate of 
  evolution, for instance on the giant brach or for clump stars or
  stars at the turnoff point, and diminishes the importance of the
  main sequence stars in the reconstruction. }
\begin{figure*}
\centering
\includegraphics[width=6cm]{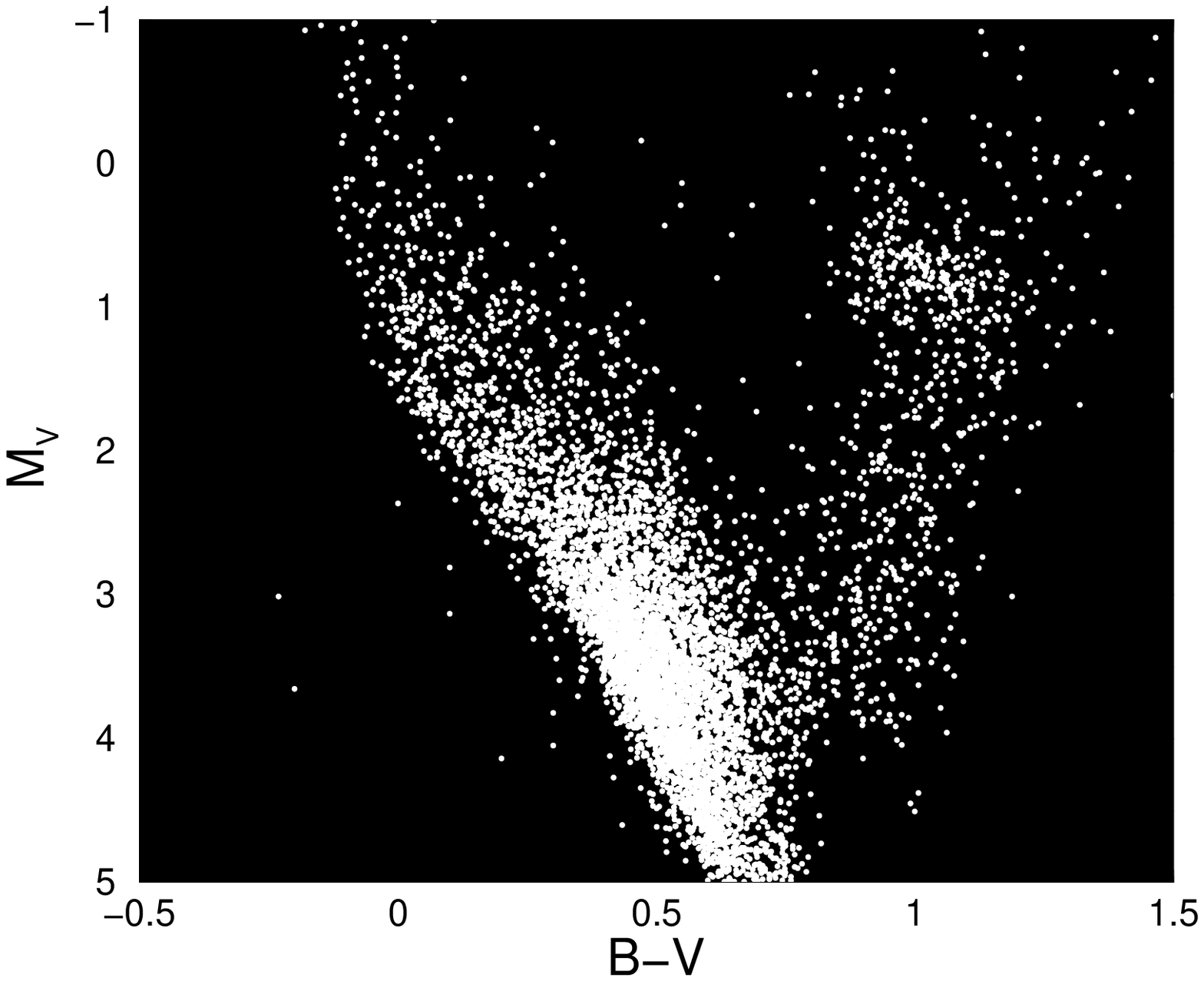}
\includegraphics[width=6cm]{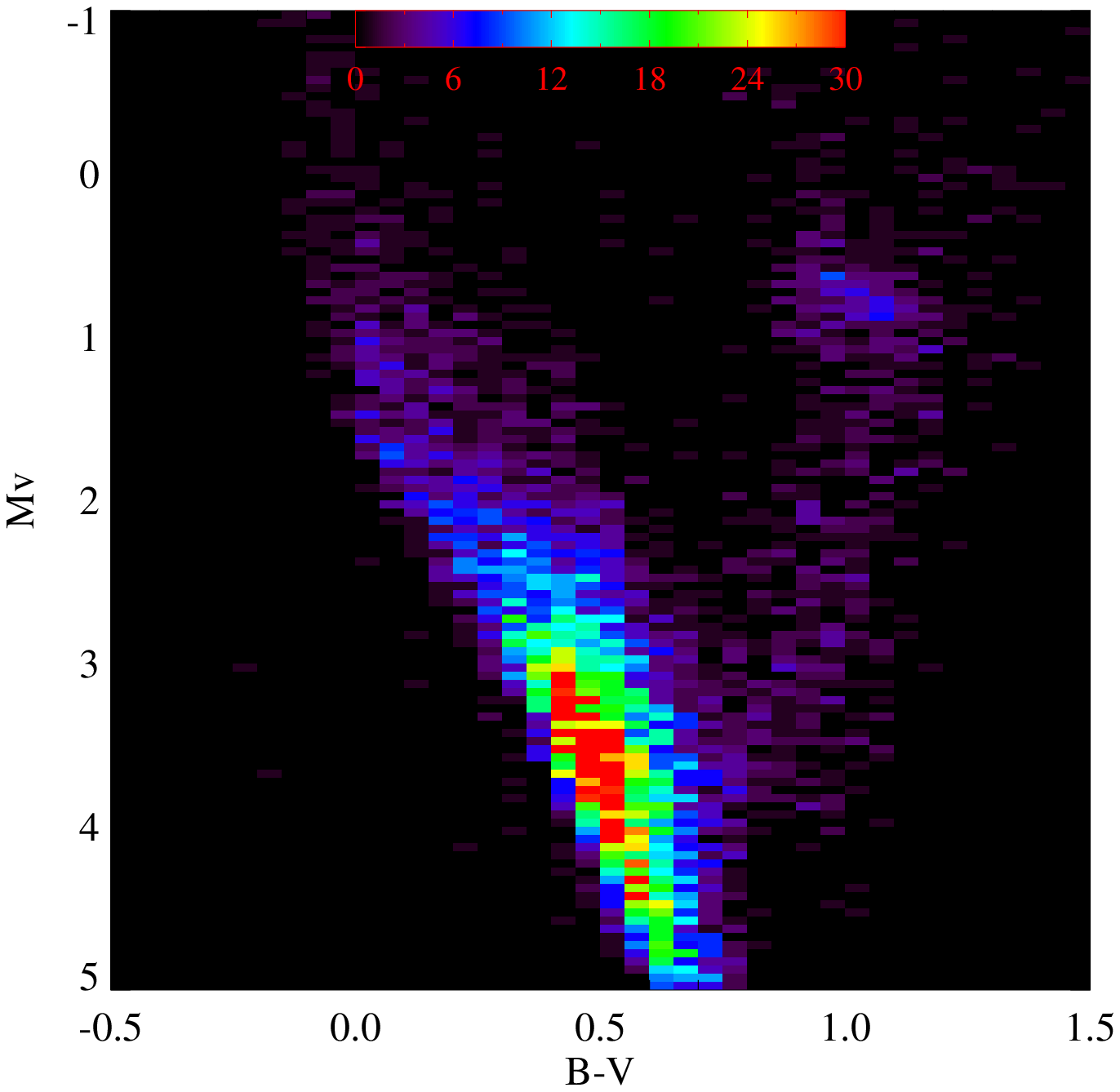}\\
\includegraphics[width=6cm]{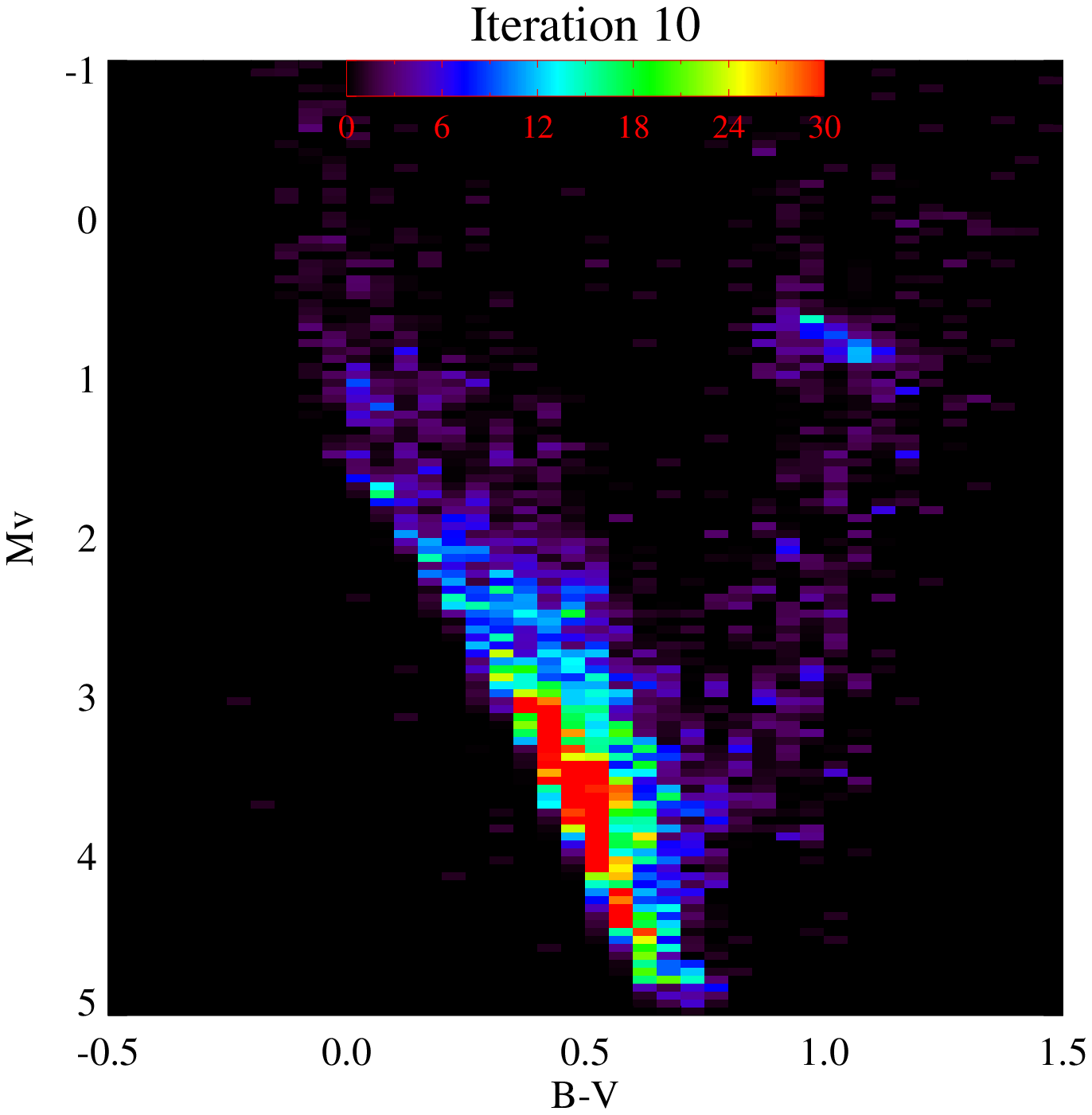}
\includegraphics[width=6cm]{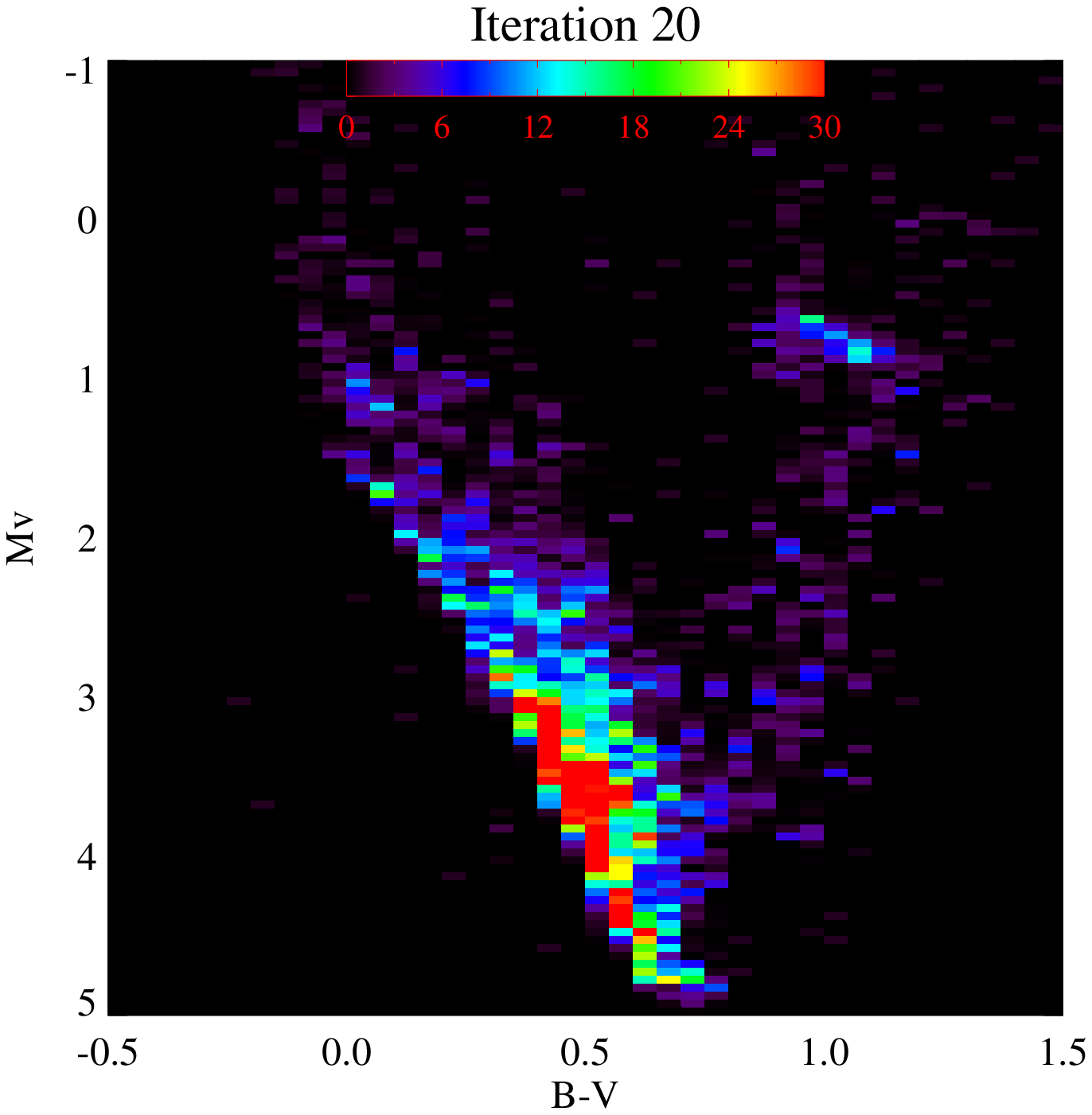}
\caption{(a) Hipparcos CMD. (b) 2-D histogram for the Hipparcos CMD. (c)-(d)
  Restorations with the pseudo-gaussian {\it psf} (the number of
  iteration is labeled). } 
\label{nmv11} 
\end{figure*}
\section{Conclusions}
The aim of this study has been to demonstrate a method for 
recovering as much information as possible from binned color-magnitude
diagrams that can then serve as input 
for evolutionary studies of stellar ensambles.  The resulting reconstructions
should be the best cleaned data set with which to perform
analyses of the star formation rates and metallicity evolution over
time for complex samples.  The advantage of this approach over
bootstrapping comes from the ease of including in the
likelihood function a broad range of processes (known or hypothetical)
that affect the location of stars within the observed CMD.  Although
we have concentrated here on a particular algorithm, the
Richardson-Lucy method, there is in principle no restriction and
others, e.g. maximum entropy, could be used instead. A study of the star formation history of
the solar neighborhood will be the topic of a future paper (Cignoni et
al. 2006, in preparation).

\begin{acknowledgements}
We thank Jason Aufdenberg, Mario Bertero, Patrizia Boccacci, 
Scilla Degl'Innocenti, Pier Giorgio Prada 
Moroni, Vincenzo Ripepi, and Carlo  Ungarelli for continuing
discussions and suggestions and an anonymous referee for {\it very} 
helpful nudging and insightful comments.  Financial support for this work was
provided by MIUR-COFIN 2003-2004.
\end{acknowledgements}

\end{document}